\documentclass[%
 reprint,
superscriptaddress,
showpacs,preprintnumbers,
 amsmath,amssymb,
 aps,
]{revtex4-1}

\usepackage{graphicx}
\usepackage{dcolumn}
\usepackage{bm}

\begin{document}
\newcommand {\be}{\begin{equation}}
\newcommand {\ee}{\end{equation}}
\newcommand {\bea}{\begin{array}}
\newcommand {\cl}{\centerline}
\newcommand {\eea}{\end{array}}
\newcommand {\pa}{\partial}
\newcommand {\al}{\alpha}
\newcommand {\de}{\delta}
\newcommand {\ta}{\tau}
\newcommand {\ga}{\gamma}
\newcommand {\ep}{\epsilon}
\newcommand {\si}{\sigma}
\newcommand{\up}{\uparrow}
\newcommand{\down}{\downarrow}

\title{Excitons and optical spectra of phosphorene nanoribbons}

\author{Zahra Nourbakhsh}
\affiliation{School of Physics, Institute
for Research in Fundamental Sciences (IPM), Tehran 19395-5531,
Iran}
\author{Reza Asgari}
\affiliation{School of Physics, Institute
for Research in Fundamental Sciences (IPM), Tehran 19395-5531,
Iran}
\affiliation{School of Nano Science, Institute
for Research in Fundamental Sciences (IPM), Tehran 19395-5531,
Iran}

\date{\today}



\begin{abstract}
On the basis of many-body {\it ab-initio} calculations, using single-shot G$_0$W$_0$ method and Bethe-Salpeter equation,
 we study phosphorene nanoribbons (PNRs) in the two typical zigzag and armchair directions.
 The electronic structure, optical absorption, electron-hole (exciton) binding energy, exciton exchange splitting,
 and exciton wave functions are calculated for different size of PNRs.
 The typically strong splitting between singlet and triplet excitonic states make PNRs favorable systems for application in optoelectronic.
 Quantum confinement occurs in both kinds of PNRs, and it is stronger in the zPNRs,
 as behave like quasi-zero-dimensional systems.
 Scaling laws are investigated for the size-dependent behaviors of PNRs.
 The first bright excitonic state in PNRs is explored in detail.
\end{abstract}
\maketitle

\section{Introduction}\label{sec:intro}

\begin{figure}
\centering
\includegraphics[width=1\linewidth] {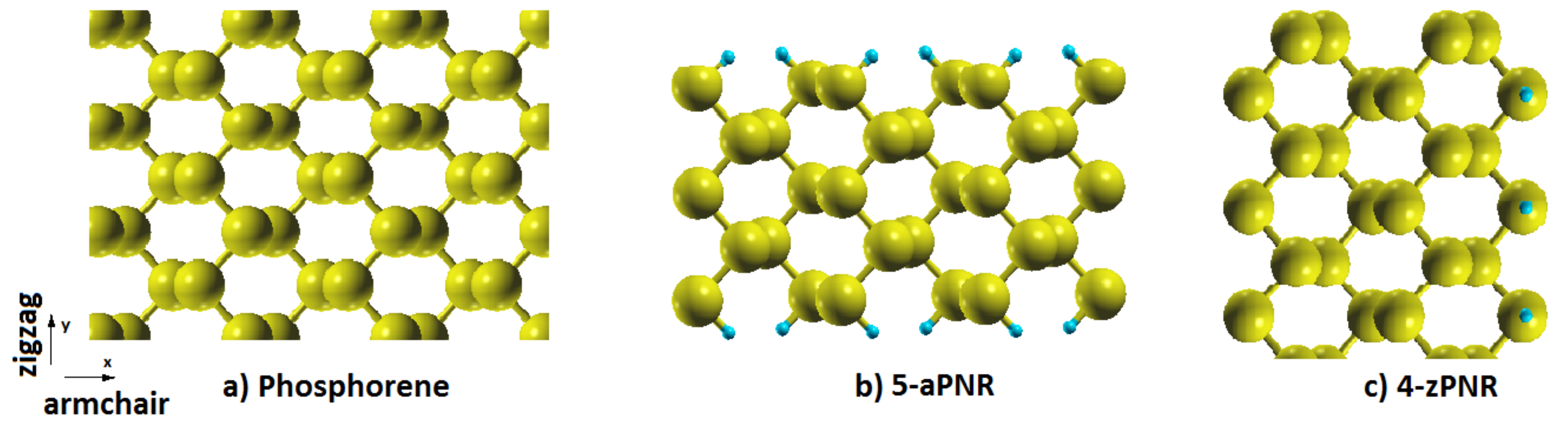}
\caption{\label{fig:structure} (Color online)
 Top view of phosphorene (a) and hydrogen terminated phosphorene nanoribbons along the armchair (b) and zigzag (c) directions. Phosphorous atoms are in yellow and hydrogen atoms (the smaller balls) are in blue. All structures are fully relaxed. In zPNRs, after relaxation, hydrogen atoms are located at top and bottom of the edge atoms with distance of $1.44$\AA. There are five (four) pairs of phosphorous atoms in the width of the aPNR (zPNR). Hence, we name these ribbons the 5-aPNRs (with width $0.86$ nm) and 4-zPNRs (with width $0.88$ nm), respectively.
 }
 \end{figure}

Since the birth of phosphorene, and its successful exfoliation from black phosphorous in  $2014$ \cite{ph-birth}, many attentions have been attracted to the unique properties of this new member of advanced two-dimensional (2D) crystalline materials. Phosphorene has a $\sim1.4~eV$ direct band gap, high carrier mobility and high on/off ratio for field-effect transistor applications \cite{ph-birth, ph-appl1, ph-appl2, ph-appl3, aniso1} beside, the fact which characterizes phosphorene as an unusual 2D system is its high anisotropic properties. In phosphorene, which is an elemental two dimensional material, each phosphorous atom from the fifth group of the  periodic table, covalently bonded to three other $P$ atoms, creating a $sp^3$ hybridization and forming a puckered honeycomb structure (shown in Fig.~\ref{fig:structure}) with a high in-plane structural anisotropy. This anisotropy has been effected in other physical behaviors of phosphorene such as electronic structure, optical absorption, transport, photonic and thermoelectronic properties \cite{aniso1, aniso2, PExcExp}. Owing to this high anisotropy, phosphorene behaves like an efficiently one-dimensional system \cite{ph-GW}.

Furthermore, the phosphorene unique anisotropic structure makes the particular behavior of the layer-dependent photoluminescence~\cite{pl} and quasi-one dimensional
excitons~\cite{ph-GW,excitons,ex}. This leads to a unique 2D platform for investigation of the dynamics of excitons in nanoribbon structure together with the many-body impacts. However,
the quasi-one dimensional excitonic nature of monolayer phosphorene can limit its photoluminescence~\cite{excitons}.

In this article, on the basis of a first-principles simulations, we are interested in studying the electronic properties of phosphorene nanoribbons which can be useful for a wide range of applications such as electronic, thermoelectric, photonic, photovoltaic, and biology. We consider different size of PNRs in two typical armchair and zigzag directions. In our simulations, the dangling bonds of PNRs are passivated using hydrogen atoms. Fig.~\ref{fig:structure} shows two structures of our studied PNR structures. Investigating the quality of occurrence of quantum confinement (QC) in PNRs, and anisotropic behaviors of optical properties, in transition from 2D to one-dimensional system, are our goals in this study.

We should note that the reduction of the system size causes a decrease of the screening and enhancement of the Coulomb interaction between charge carriers (electrons and holes in  semiconductors). Therefore, the excitonic effects are no longer negligible in nanostructures and the single particle Kohn-Sham density functional theory (DFT) is not reliable to study the electronic and optical properties of these systems. Accordingly, we do have to enter the many-body interactions in our simulation going beyond of standard DFT and accept the cost of heavier calculations.

We carry out an {\it ab-initio} electronic calculations, using single-shot G$_0$W$_0$ method~\cite{dftgap,gw} and Bethe-Salpeter~\cite{bse} equation in order to include many-body interactions.
Within this approach, we find the quasiparticle (QP) band gap and exciton binding energy of phosphorene in a very excellent agreement with experiment \cite{PExcExp}.
Also results obtained by this method are in good agreement with those obtained in experiments for one-dimensional systems such as silicon nanoribbons \cite{Sinw}.

The paper is organized as follows. The theoretical and technical method is discussed in Sec. II and then
we present our results in Sec. \ref{sec:results}.
Introducing the studied structures,
QP and optical band gap, investigating the scaling law of the band gap tunability by the ribbon size,
exploring the excitonic effect and exchange splitting of excitons in the studied systems,
analysing the optical spectra, and
displaying the excitonic wave function in a ribbon and 2D structures of phosphorene,
are presented in this Section. Finally, we summarize and conclude our main results in Sec. \ref{sec:conclusion}.

\section{Theory and computational details}\label{sec:method}

The first-principles simulations based on density functional theory are carried out by using the PWscf code of the Quantum Espresso package \cite{QE}. The generalized gradient approximation (GGA) in the scheme of Perdew, Burke, and Ernzerhof (PBE) \cite{pbe} and norm-conserving pseudo-potentials are used throughout our calculations. The Kohn-Sham single particle wave functions are expanded in plane-waves up to an energy cutoff of $75$~Ry and the Fourier expansion of electron density are cut at $300$~Ry. The Brillouin zone integrations are performed on the Monkhorst-Pack \cite{m-p} k-point grid of $16 \times 12 \times 1$ for 2D phosphorene. To avoid the unphysical interactions between periodic images and simulate the isolated system, we use the vacuum layers of $25$\AA. The geometrical optimization is performed according to the force and stress and it is satisfied when all components of all forces are less than $10^{-3}$~eV/\AA.

Here, after geometrical relaxation on the basis of DFT-GGA, we perform three steps:
the first step is calculating the ground-state Kohn-Sham wave functions \cite{k-sh} from the equation below
\begin{equation}
[-\frac{\hbar^2}{2m}\nabla^2+V_{ion}(r)+V_{Hartree}(r)+V_{xc}(r)]\varphi_i=\varepsilon_i\varphi_i
\end{equation}
where $V_{ion}(r)$ is the potential felt by the electrons produced by the ions in their equilibrium
positions, $V_{Hartree}(r)$ is the Hartree part of the Coulomb electron-electron interaction and $V_{xc}(r)$ is the many-body exchange-correlation potential. The self-energy of quasiparticles, $\Sigma(E)$ is calculated using single-shot G$_0$W$_0$ approximation \cite{gw} and therefore, the QP corrections to the GGA eigenvalues are estimated from
\begin{equation}\label{energy}
E_i=\varepsilon_i+ \langle\varphi_i|\Sigma(E_i)-V_{xc}|\varphi_i\rangle
\end{equation}

The BSE can be solved separately for spin singlet
and spin triplet excitons when the spin-orbit term is negligible. We can write the Bethe-Salpeter Hamiltonian in the Fock space of the pairing ($|eh\rangle$) and anti-pairing $|\widetilde{he}\rangle$ of electron-hole \cite{bse2, bse3} in the form of
\begin{equation}
H=
\begin{bmatrix}\label{h1}
 R  & C\\
 -C^* & -R^*
\end{bmatrix}
\end{equation}
where its eigenfunction has a form of
$\sum_{eh}{ A^S_{eh} ~ |eh\rangle + B^S_{\widetilde{he}} ~ |\widetilde{he}\rangle }$ where
$A^{S}$ and $B^{S}$ are the exciton amplitudes of the state $S$ derived from solving the BSE.
The diagonal (Resonating) and anti-diagonal (Coupling) terms in the Bethe-Salpeter Hamiltonian are
\begin{equation}
R_{eh;{e'}{h'}}=(E_h-E_e)\delta_{e{e'}} \delta_{h{h'}}+ \langle eh|K|{e'}{h'}\rangle
\end{equation}
and
$C_{eh;{e'}{h'}}=\langle eh|K|\widetilde{{h'}{e'}}\rangle$
where $E_e$ and $E_h$ are the electron and hole energies given by Eq.~\ref{energy}, respectively, and $K$ is the four point BSE interaction kernel, and in principle contains all possible interactions between two particles. The kernel $K$ has two terms; $K=K^d+K^x$ in which the first term is called the direct term, describes the screened electron-hole attractive interaction that is responsible for the creation of the excitons and the second term is the exchange repulsive electron-hole interaction and results from the bare Coulomb interaction. In the spin triplet excitons case, the exchange term is zero and therefore, the triplet exciton is lower in energy owing to the lack of the repulsive exchange term.

The off-diagonal blocks in the Bethe-Salpeter Hamiltonian couple the electron-hole pairing and anti-pairing and they usually have a very small and ignorable value.  This assumption is known as the Tamm-Dancoff approximation (TDA)\cite{tda}. The coupling terms are non-Hermitian, thus TDA converts a non-Hermitian problem to a Hermitian one.

On the basis of DFT-PBE, the lattice parameters of 2D phosphorene are obtained $a_{ac}=4.6$\AA~and $a_{zz}=3.3$\AA~which are in good agreement with those values reported by experiment \cite{explat}.
Also its band gap is $0.96$~eV which is close to other DFT-PBE calculations~\cite{asgari}.

\begin{figure}
\centering
\includegraphics[width=1.0\linewidth] {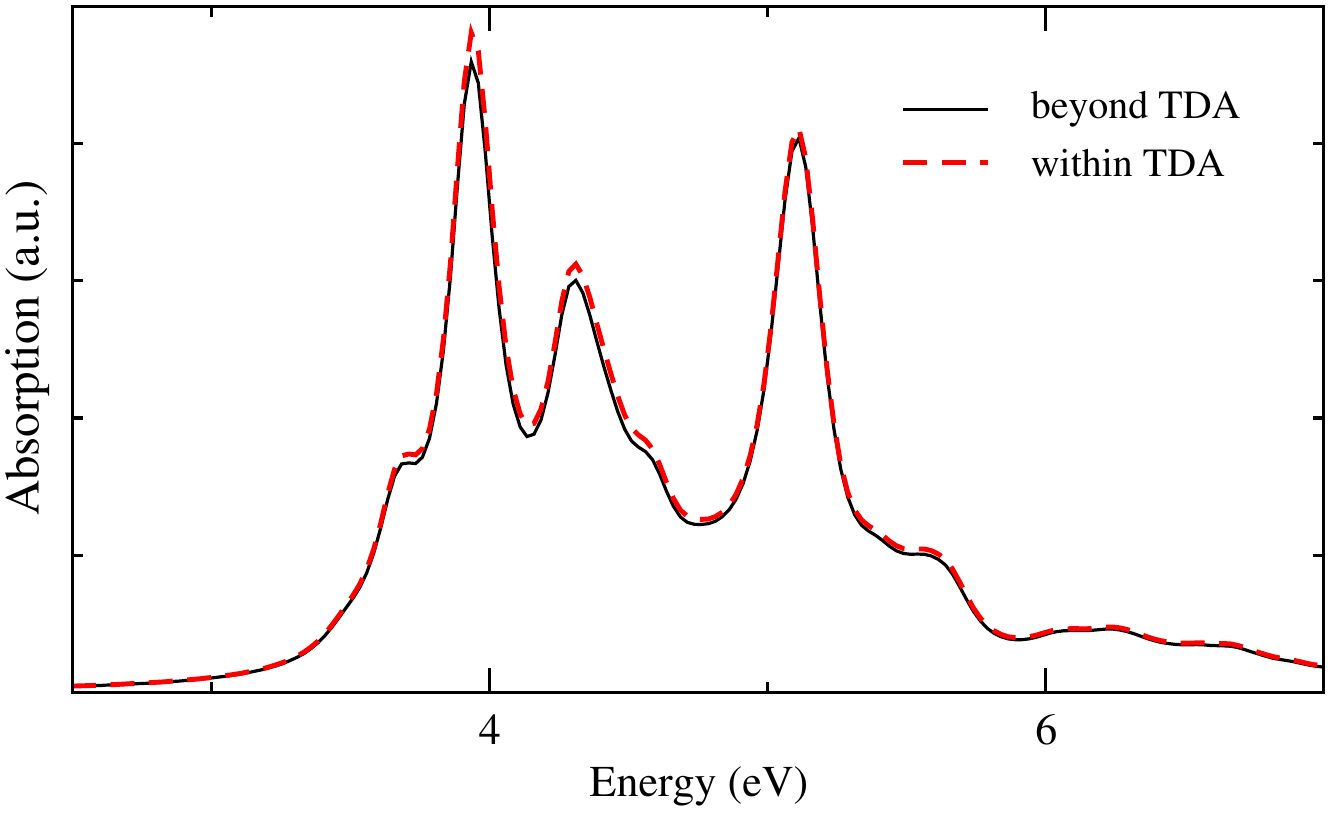}
\caption{\label{fig:TDA} (Color online) Optical spectra of the 4-zPNR within and beyond Tamm-Dancoff approximation. Notice that the difference between the two results is very small. }
\end{figure}

All of the methods beyond DFT calculations incorporating the many-body effects through the G$_0$W$_0$ method and BSE are performed using the YAMBO \cite{yambo} code.
The QP energy is calculated by the single-shot G$_0$W$_0$ approximation with the general
Coulomb hole plus screened exchange (COHSEX) scheme \cite{cohsex}.
Excitonic interactions, and optical spectra are obtained by solving the BSE.
We check the accuracy of TDA in our systems.
The absorption spectra of the four pairs phosphorous atoms along the zigzag nanoribbon (4-zPNR) within and beyond TDA are presented in Fig. \ref{fig:TDA}.
It shows that the effect of the coupling blocks in the Bethe-Salpeter Hamiltonian are negligible in absorption. As shown, there is no evidence of the first peak,  which is dark within the single photon process owing to the symmetry of the system. Since the discrepancy between the two results is tiny, it shows that the off-diagonal term in Eq.~\ref{h1} does not play important role and thus we can safely use TDA approach in our numerical results.

The long-range nature of the Coulomb interaction, in both G$_0$W$_0$ and BSE calculations, induces improper interactions between periodic image. To avoid this deviation, the Coulomb interaction is truncated with the box-shape with the size of at least $22$\AA~in each direction. To attain the accuracy of $0.1$~eV in the calculation of the QP band gaps, excitonic energies, and optical spectra, we check the convergency with respect to different parameters. To calculate the polarization function, the number of unoccupied band is at least four times of occupied states. The beyond DFT calculations are done in the dense $k$-point grid of  $35\times 50\times 1$ for 2D phosphorene, and $40\times 1\times 1$ and $50\times 1\times 1$ for armchair PNRs (aPNRs) and zPNRs, respectively.

\section{Numerical Results}\label{sec:results}

Using DFT-PBE, Tran and Yang have calculated the band gap and optical absorption of PNRs \cite{pnrs}.
Here, we are interested in exploring the many-body effects on these properties.
We consider PNRs, oriented along the $ac$ and $zz$ directions, with
different widths of $5$ ($8.8$) up to $18.6$ ($27$) angstrom in aPNRs (zPNRs).

The PNR formation energy, $E_f$, is calculated as $E_f=E^{PNR}_0(n,m)-(nE_0^{ph}+\frac{m}{2}E_0^{H_2})$ where $E^{PNR}_0(n,m)$ denotes the ground-state energy of the PNR, included $n$ phosphorous atoms and $m$ hydrogen atoms, $E_0^{ph}$ and $E_0^{H_2}$ are ground-state energies of 2D phosphorene per atoms and hydrogen molecule, respectively. A negative value of $E_f$ means that energy is released by forming PNR. On the contrary, the positive value indicates endothermic formation of the PNR. In nanostructures, the formation energy is usually positive; it means that we have to work on the system to create the nanostructure.

The calculated formation energies of PNRs have a very small (several $10$~meV)
positive (negative) values for aPNRs (zPNRs). It reveals the high tendency of phosphorene
to combine with hydrogen and in consequences its instability in the ambient conditions \cite{ph-instability1,ph-instability2}.
For the sake of comparison, we calculate the formation energy of the hydrogen-terminated graphene nanoribbons and $E_f$ is about $1.3-1.4$~eV in the case of armchair or zigzag gaphene nanostructures .
Also, in all the edge free PNRs, the formation energies
have a positive value in the range of several electron-volts.

\subsection{Band Gap and Scaling Laws in PNRs}\label{sec:gap}

\begin{figure}
\centering
\includegraphics[width=1.0\linewidth] {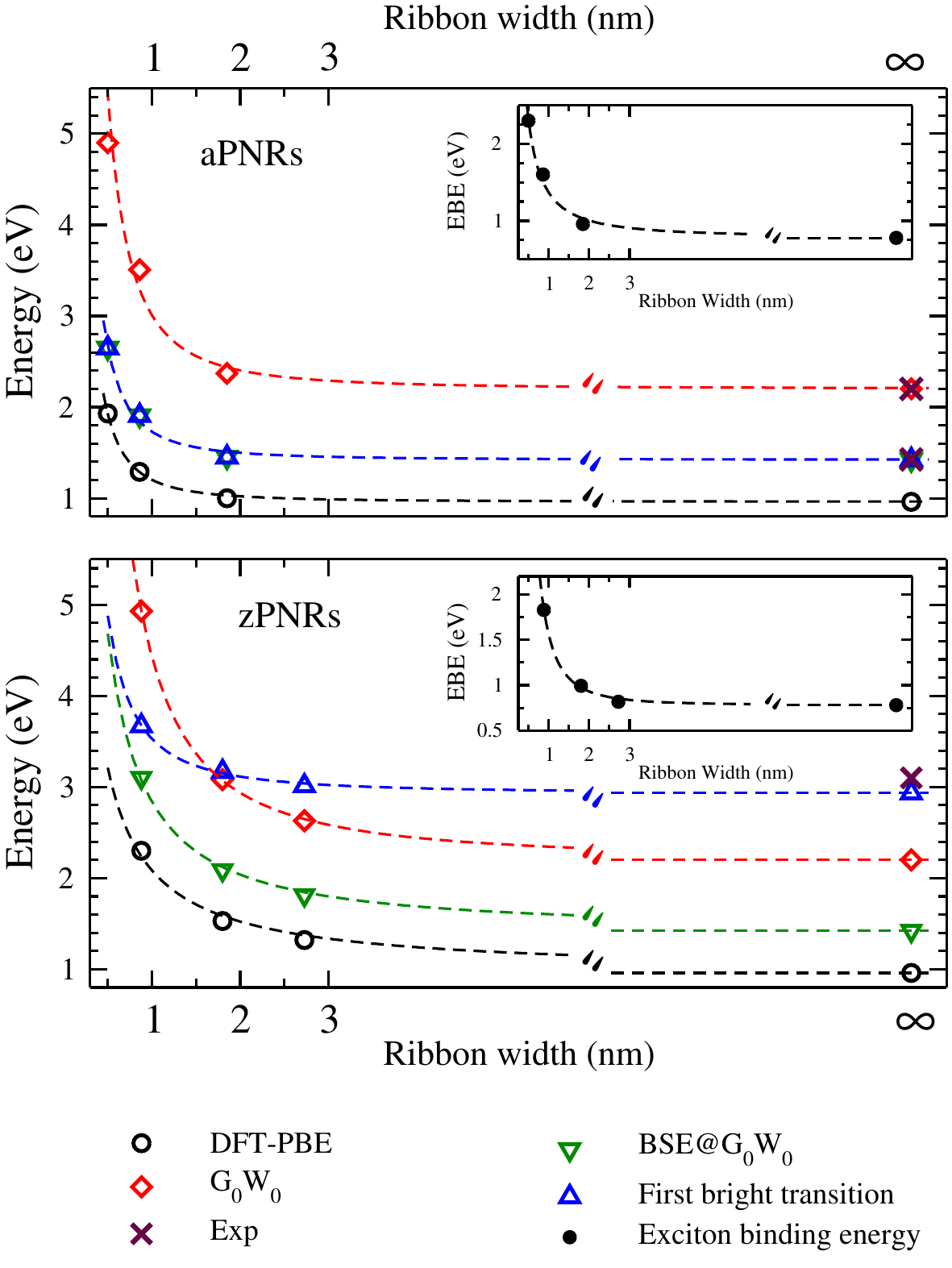}
\caption{\label{fig:gap} (Color online)
The evolution of the band gaps (PBE, G$_0$W$_0$), optical gap (BSE$@G_0W_0$) and first bright excitonic state as a function of the nanoribbon width.
The insets show the variations of the exciton binding energy (EBE) with the ribbon width.
The power law fitting curves are presented by dashed lines. Experimental data of 2D phosphorene are marked by the cross sign \cite{PExcExp,exp-p}.
}
\end{figure}

\begin{table}
\caption{\label{table:gap}
Fitted parameters of the DFT gap (PBE), QP gap (G$_0$W$_0$), optical gap (BSE$@G_0W_0$), and exciton binding energy (EBE) as a function of ribbon widths by the formula $\Delta=A/w^{\alpha}+E_{w\rightarrow\infty}$. FE and FBE denote to the first excitonic bound state and first bright excitonic bound state. All the diagrams are plotted in Fig. \ref{fig:gap}.}
\begin{center}
 \begin{tabular}{ c c c c c c c}
 \hline
 PNRs &  & DFT    &   QP  &\multicolumn{2}{c}{Optical Gap}&~~EBE~~\\
      &	  &~~Gap~~&~~Gap~~& FE & FBE  \\
\hline
ac     & A & 0.24  & 0.82  & 0.31  & 0.31  & 0.59 \\
	& $\alpha$ & 2.0 & 2.0 & 2.0 & 2.0 & 1.39 \\
	& $E_{w\rightarrow\infty}$  & 0.96 & 2.2 & 1.42 & 1.42 & 0.78 \\
 \hline
zz     & A & 1.13  & 2.23  & 1.42  & 0.60  & 0.78 \\
	& $\alpha$ & 1.0 & 1.60 & 1.29 & 1.70 & 2.37 \\
	& $E_{w\rightarrow\infty}$ & 0.96 & 2.2 & 1.42 & 2.97 & 0.78 \\
\hline
\multicolumn{2}{c}{2D phosphorene} & Exp\cite{PExcExp} & 2.2$\pm0.1$ &$1.3\pm0.02$ & & $0.9\pm0.12$\\
		&		   & Exp\cite{exp-p} &  & 1.55 & 3.14$(zz)$ &  \\
    &\multicolumn{2}{c}{other calculation \cite{ph-2dmat}}& 2.02 & 1.24 & & 0.78 \\
\hline
\end{tabular}
\end{center}
\end{table}

Fig.~\ref{fig:gap} shows the variations of the band gap and exciton binding energy (EBE) versus the ribbon width, along in both $ac$ and $zz$ ribbon directions. As it is displayed, the calculated band gaps decrease by  increasing the ribbon width, and eventually they merge to the related values of 2D phosphorene. We fit the band gaps and EBE with a power law function of the form $A/w^{\alpha}+E_{w\rightarrow\infty}$ where $w$ is the nanoribbon width. The resulted parameters are reported in Table \ref{table:gap}. The fitting parameters in the DFT level, are in a good agreement with previous {\it ab-initio} calculations \cite{pnrs}. However, DFT (LDA or GGA) results underestimates the band gap in comparing with experiment or G$_0$W$_0$ calculations \cite{dftgap}. It is worthwhile mentioning that we find the parameter $\alpha=2$ for an armchair structure within all approaches, while it changes for a zigzag structure in each approach. The scaling parameters obtained here are also comparable with the same values of graphene ($\alpha=1$ \cite{gnr}) and silicon nanoribbons ($\alpha=1.7-2$ \cite{Si-nr2}).

The optical properties are derived from the incident light polarized along the ribbon orientation. In all the zPNRs, the first excitonic states are dark; it means that the gap transition is forbidden due to the dipole selection rules~\cite{bright} (we will discuss this point in detail in Sec.~\ref{sec:excwave}). This behavior also occurs in 2D phosphorene when the incident light polarization is along the $zz$ direction. In zPNRs, we also plot the energy gap refers to the first bright excitonic state.

In 2D phosphorene, the calculated DFT band gap, G$_0$W$_0$ band gap, optical gap, EBE, and
the energy of the first bright excitonic state (when the light polarization is along the $zz$ direction)
 are reported in Table~\ref{table:gap} as the infinity limits of PNRs behavior.
All of these values are in a very good agreement with those obtained by experiments \cite{PExcExp, exp-p} and other theoretical result \cite{ph-2dmat}.

The inset plot in Fig.~\ref{fig:gap} shows increasing of the EBE (E$_{EBE}$=E$_{G_0W_0}$-E$_{BSE@G_0W_0}$) by decreasing the ribbon size. This behavior originates from the enhancement of the overlap between electron and hole wave functions in the confined systems. Furthermore, the reduction of the screening by the size leads to increase these energies.

In Fig. \ref{fig:gap}, for the same width $ac$ and $zz$ nanoribbons,
the band gap of the zPNR is larger than the related aPNR gap value.
It means that quantum confinement in zPNRs is stronger.
In Table~\ref{table:gap}, the coefficient $A$ (the exponent $\alpha$) of aPNRs in compare with
 zPNRs is smaller (larger) and it confirms weaker QC in aPNRs.

\begin{figure}
\centering
\includegraphics[width=0.8 \linewidth] {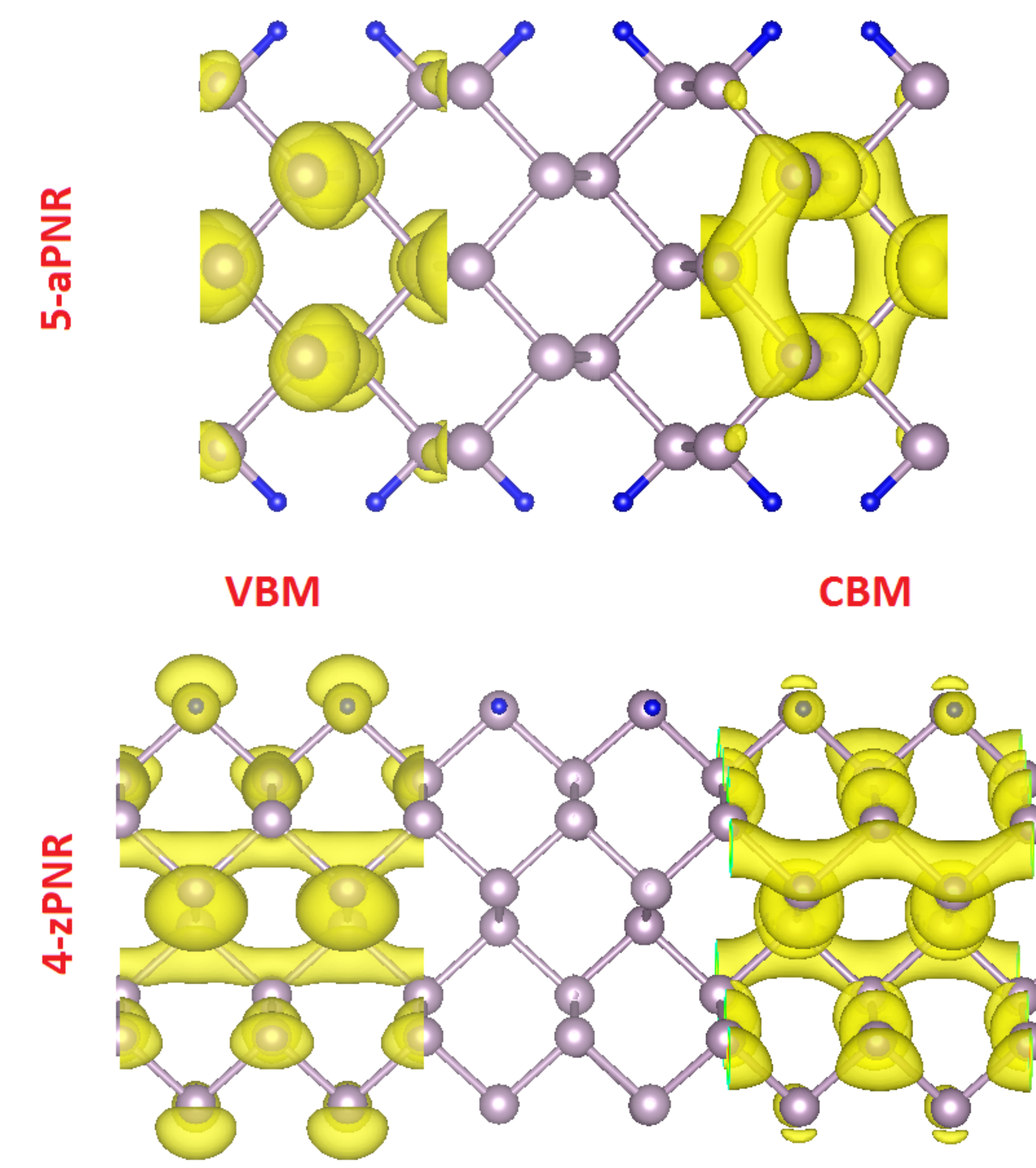}
\caption{\label{fig:hl} (Color online) Top view of electron charge density of
the VBM and CBM states of the $5-aPNR$ and $4-zPNR$.}
\end{figure}

Fig.~\ref{fig:hl} shows the distribution of the charge density of the valence band maximum (VBM) and conduction band minimum (CBM) states of the 5-aPNR and 4-zPNR as an example of aPNRs and zPNRs. It shows that in zPNR, the CBM and VBM states are distributed in the ribbon width. While in the aPNR, these states are strong in the middle. So they do not feel the edge effects as much as zPNRs. Consequently, the confinement in aPNRs is weaker. As we discuss in continuing  this anisotropic behavior can be explained by the anisotropy of the effective masses in phosphorene.

\begin{figure}
\centering
\includegraphics[width=1.0\linewidth] {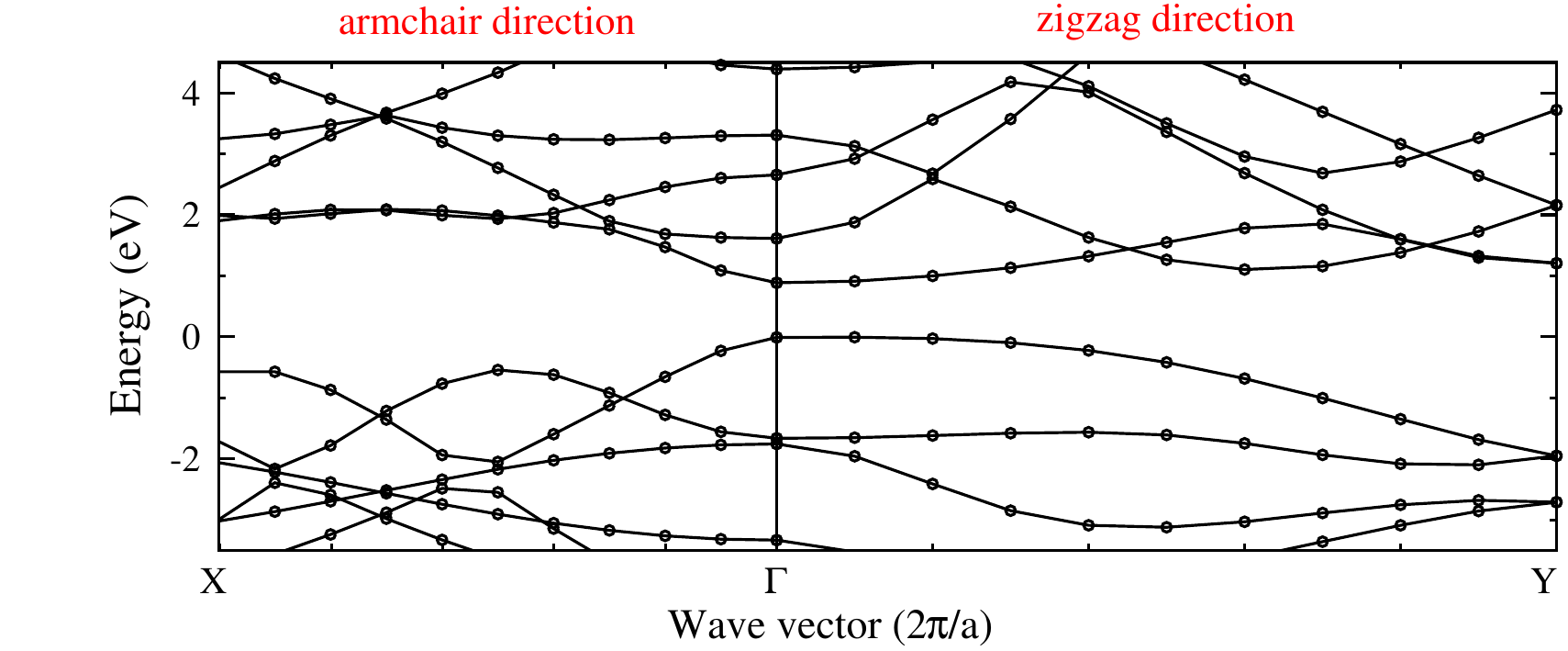}
\caption{\label{fig:pband} (Color online) Electronic band structure of 2D phosphorene, calculated by DFT-PBE along the $ac$ and $zz$ directions.
 The top of the valence band is set to be zero.
}
\end{figure}

The QC happens when the system size becomes less than exciton Bohr radius of the bulk system.
This radius is in proportion to the inverse of the reduced exciton effective mass,
$a^*_B \propto \frac{1}{\mu_{ex}} = \frac{m_e + m_h}{m_e \times m_h}$ \cite{grosso}.
The effective masses of carriers in phosphorene are strongly anisotropic \cite{exp-p}.
Fig.~\ref{fig:pband} shows the band structure of 2D phosphorene.
Around the gap (at $\Gamma$ point), the band structure is highly flat along the $zz$ direction
and nearly linear along the $ac$ direction~\cite{asgari}.
It indicates that carriers in phosphorene behave like relativistic light particles
along the armchair direction while they are classical heavy particles along the zigzag direction. Therefore,
\begin{equation}
\mu_{ex}^{ac}  \textless~ \mu_{ex}^{ zz} \Rightarrow a^*_{B-zz} \textless~ a^*_{B-ac}
\end{equation}
accordingly, QC in aPNRs, which are confined along the $zz$ direction, shows less widths in compare with zPNRs.

\begin{figure}
\centering
\includegraphics[width=0.9\linewidth] {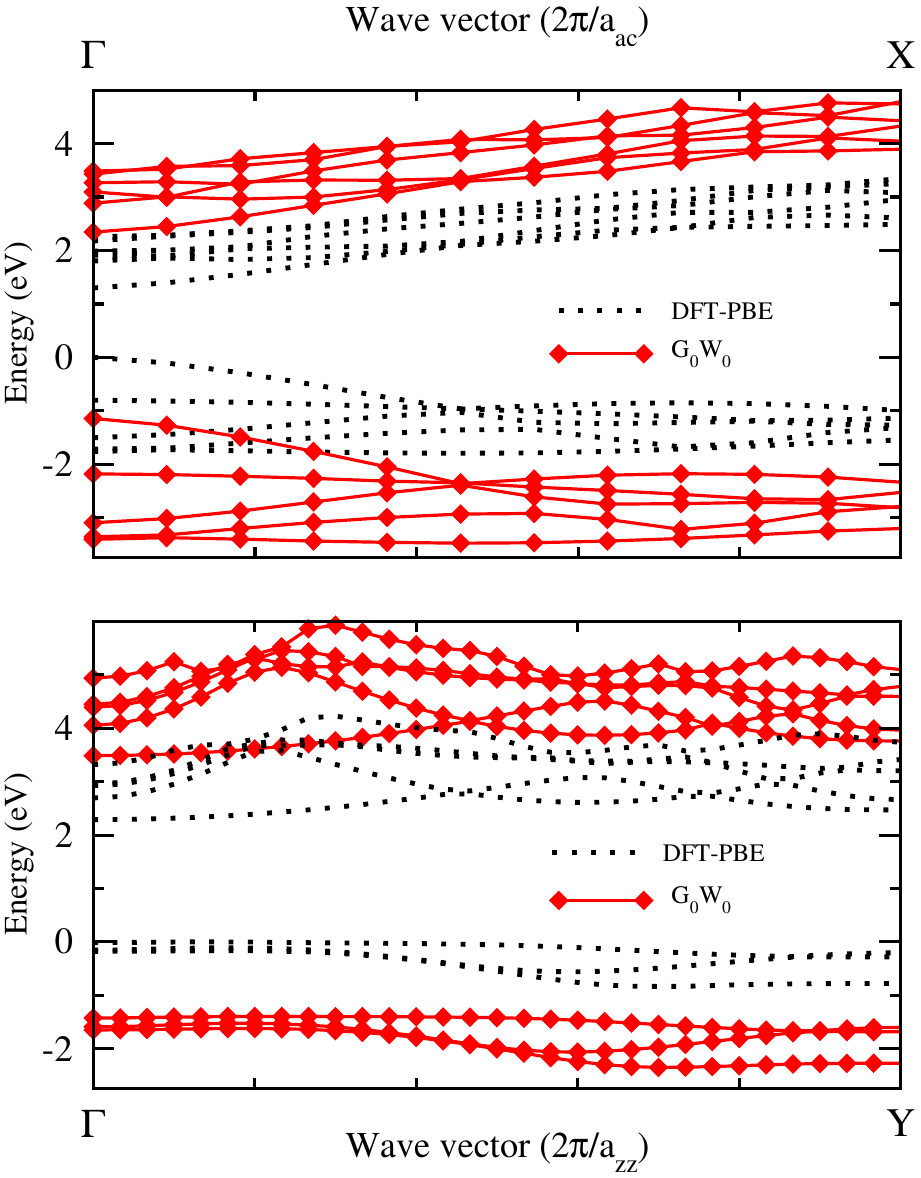}
\caption{\label{fig:pnrBand}
(Color online) KS (PBE) and QP (G$_0$W$_0$) band structures of the 5-aPNR (top panel) and 4-zPNR (bottom panel).
Both PNRs have a direct band gap at $\Gamma$ point.
For 4-zPNR, the band structure is very flat and it behaves like a zero-dimensional system. The PBE-VBM states are set to be zero.
}
\end{figure}

Fig.~\ref{fig:pnrBand} shows the band structure of the 5-aPNR and 4-zPNR in the DFT and G$_0$W$_0$ level. The band gap of both structures is direct and located at $\Gamma$ point. At DFT level, the valence band maximum is not exactly on $\Gamma$ point in zPNRs, however, their difference is very small ($\sim0.01~eV$) and we can consider a direct band gap for zPNRs. In the G$_0$W$_0$ level, their difference is much less than our numerical accuracy. Fig.~\ref{fig:pnrBand} shows the QP corrections slightly increase the bandwidth owing to the increasing the kinetic energy of the QPs. Notice that the PNRs with the edge phosphorous atoms passivated by hydrogen are direct gap semiconductors, however, the pristine zPNRs are metals for all values of the ribbon width, while the pristine aPNRs are semiconductors with indirect band gaps \cite{pnrs, pnrs2}.

\begin{figure}
\centering
\includegraphics[width=1.0\linewidth] {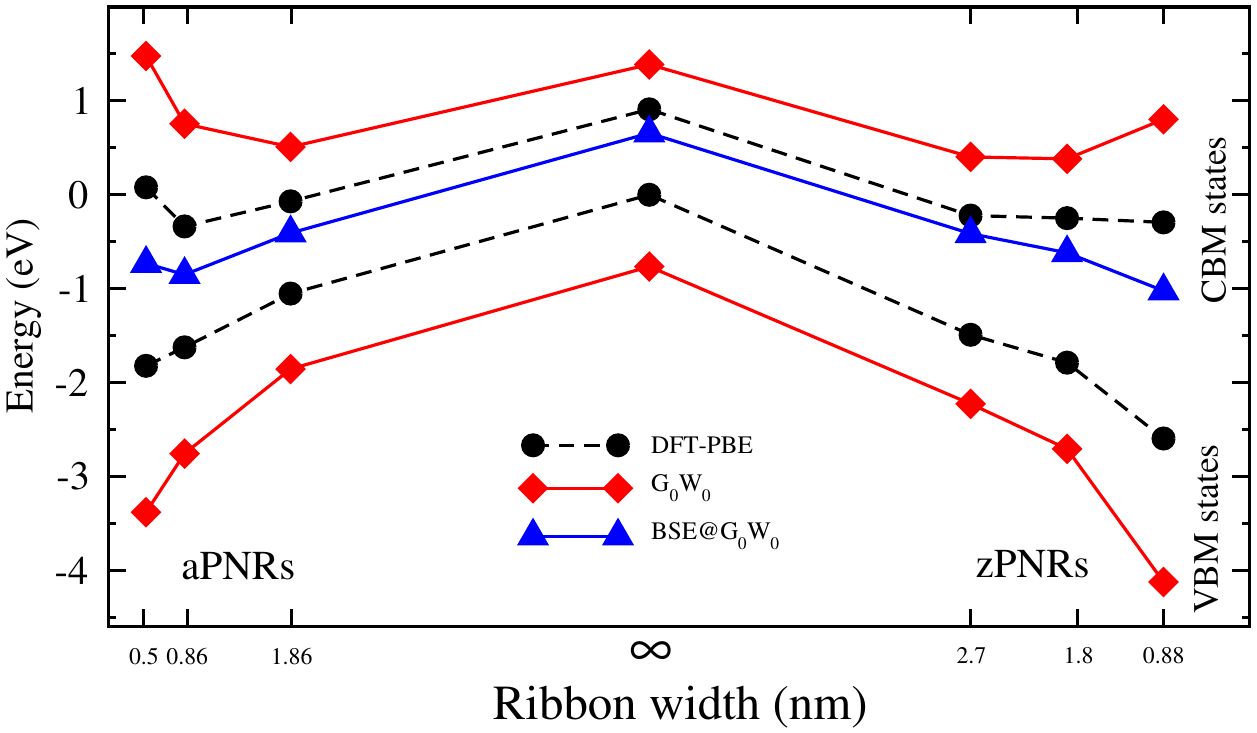}
\caption{\label{fig:vbmcbm} (Color online)
The calculated VBM and CBM states of PNRs and 2D phosphorene as a function of their width.
The G$_0$W$_0$ correction on the PBE states and the excitonic effects on the G$_0$W$_0$ states are specified. The BSE$@$G$_0$W$_0$ is evaluated from the VBM energy obtained by G$_0$W$_0$ approach. The PBE-VBM state of phosphorene is set to be zero and all data are shifted accordingly.}
\end{figure}

In Fig.~\ref{fig:vbmcbm}, we display the evolution of the energy of the VBM and CBM  states of PNRs as function of their sizes. The  energy of both the VBM and CBM states increased by increasing the size meaning that in the PNR hetrojunction, the lowest energy state of electrons and holes are in the opposite side which is an important property in the photovoltaic applications.

\subsection{Excitons in PNRs}\label{sec:excEnergy}

\begin{figure}
\centering
\includegraphics[width=1.0\linewidth] {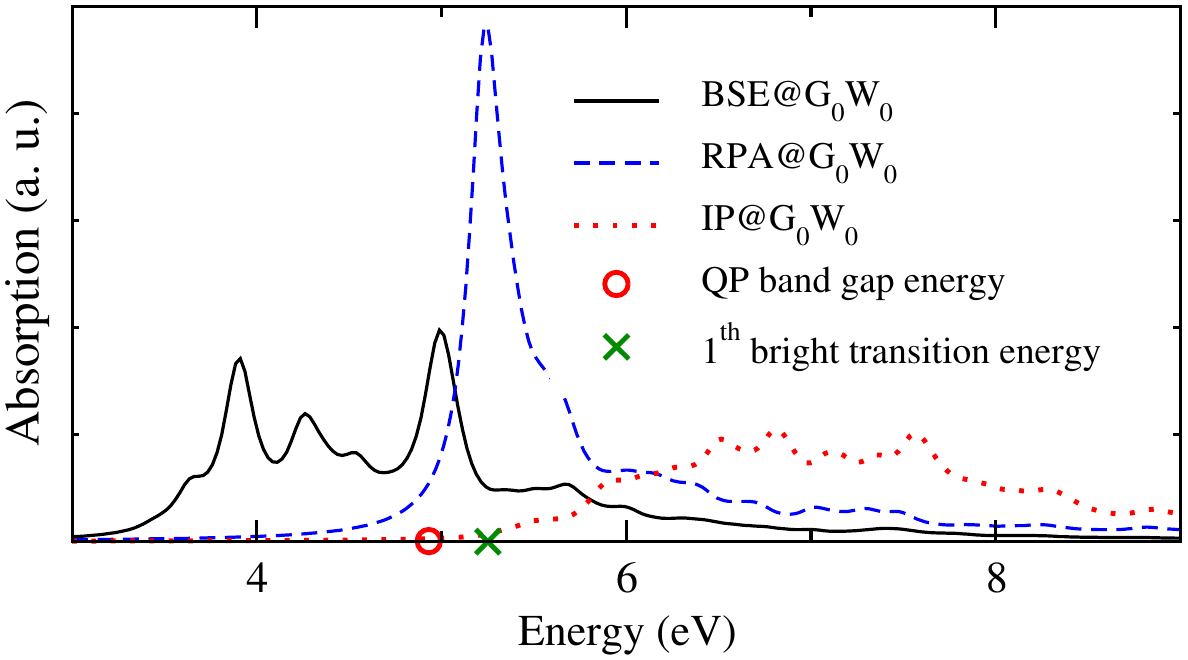}
\caption{\label{fig:abs}
(Color online) The optical absorption of the 4-zPNR for light polarization along the $zz$ direction. The BSE is reported by the solid line, RPA by a dashed line and Independent Particle (IP) by dot line. All of spectra are calculated based on the G$_0$W$_0$ quasi-particle band structure and furthermore, a peak due to a bound exciton arises at lower energy.
}
\end{figure}

In order to show the importance of many-body interactions and excitonic effect on the optical behavior of PNRs, we calculate and plot in Fig. \ref{fig:abs} the absorption spectrum calculated by the  imaginary part of the macroscopic dielectric function in the long wavelength limit of the 4-zPNR both electron-hole interaction (by solving BSE) and without electron-hole interaction (in random-phase approximation (RPA) and IP levels). It is well-known that RPA is to be inadequate for optical properties due to the electron-hole interaction~\cite{Reining}. The optical spectra clearly show that the effect of many-body interactions is not negligible. After including the electron-hole interactions, the peak of the excitonic bound states appears below the band gap with an EBE of $\sim 1.5$~eV, while RPA only gives the absorption edge of the inter band transitions. Furthermore, the strength of the peaks is completely different between BSE and RPA. Notice that the gap point transitions are forbidden in both BSE and RPA levels.

\begin{figure}
\centering
\includegraphics[width=.90\linewidth] {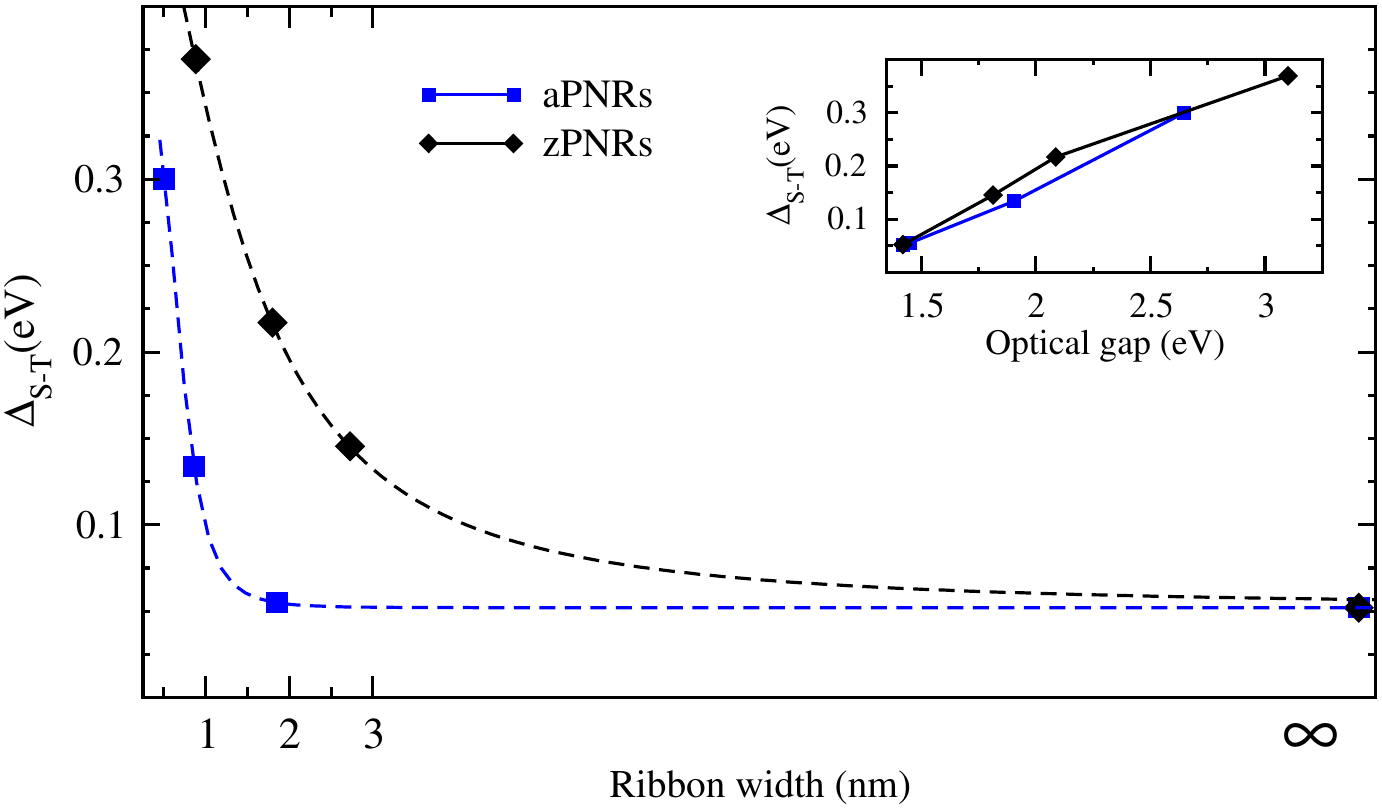}
\caption{\label{fig:delST}
(Color online) Exciton exchange splitting versus ribbon width.
In the inset, the exchange splitting as a function of the band gap of PNRs.}
\end{figure}

Fig.~\ref{fig:delST} shows the excitonic exchange splitting ($\Delta_{S-T}$ is the energy splitting between the lowest spin singlet and triplet excitonic bound states) of PNRs as a function of ribbon widths (and versus optical band gap of PNRs in the inset). According to this figure, the exchange splitting increases by decreasing the size and it is stronger for zPNRs. This is owing to the enhancement of the overlap between electron and hole wave functions by reducing the system size. It thus increases the exchange interaction in the BSE exchange-kernel. As we mentioned before (Sec. \ref{sec:method}) because of the repulsive exchange interaction term, the energy of the singlet state is always higher than the triplet state.

In 2D phophorene (infinity width limit shown in Fig. \ref{fig:delST}), the exchange splitting is $52$~meV which is in agreement with the previous calculation \cite{ph-2dmat}. This splitting energy is higher than the energy of the thermal fluctuation and also spin-orbit interaction energy of phosphorene \cite{Psoc,exp-p}. Therefore, the spin singlet and triplet could be good quantum numbers. As a consequence, the triplet excitons have a long lifetime (in the range of a mile-second to infinite if the total spin is a good quantum number). $\Delta_{S-T}$ in phosphorene and PNRs are really large in comparing with other nanostructures such as doped silicon-nanowires ($\sim100$~meV~\cite{Sinw}), graphyne ($\sim 150$~meV~\cite{graphyne}) and Carbon nanotubes ($\sim20$~meV~\cite{cnt}) which are known as materials to impressive spin singlet-triplet splitting. The giant singlet-triplet splitting are interesting in photovoltaic, biomedical, photoluminescent and quantum information applications~\cite{graphyne, ph-2dmat}.

We should also note that photon absorption can not make a spin flip in the system,
so in the triplet exciton creation, other interactions such as spin-orbit are required.

\subsection{Absorption Spectra and Excitonic wave function in PNRs}\label{sec:excwave}

\begin{figure}
\centering
\includegraphics[width=0.750\linewidth] {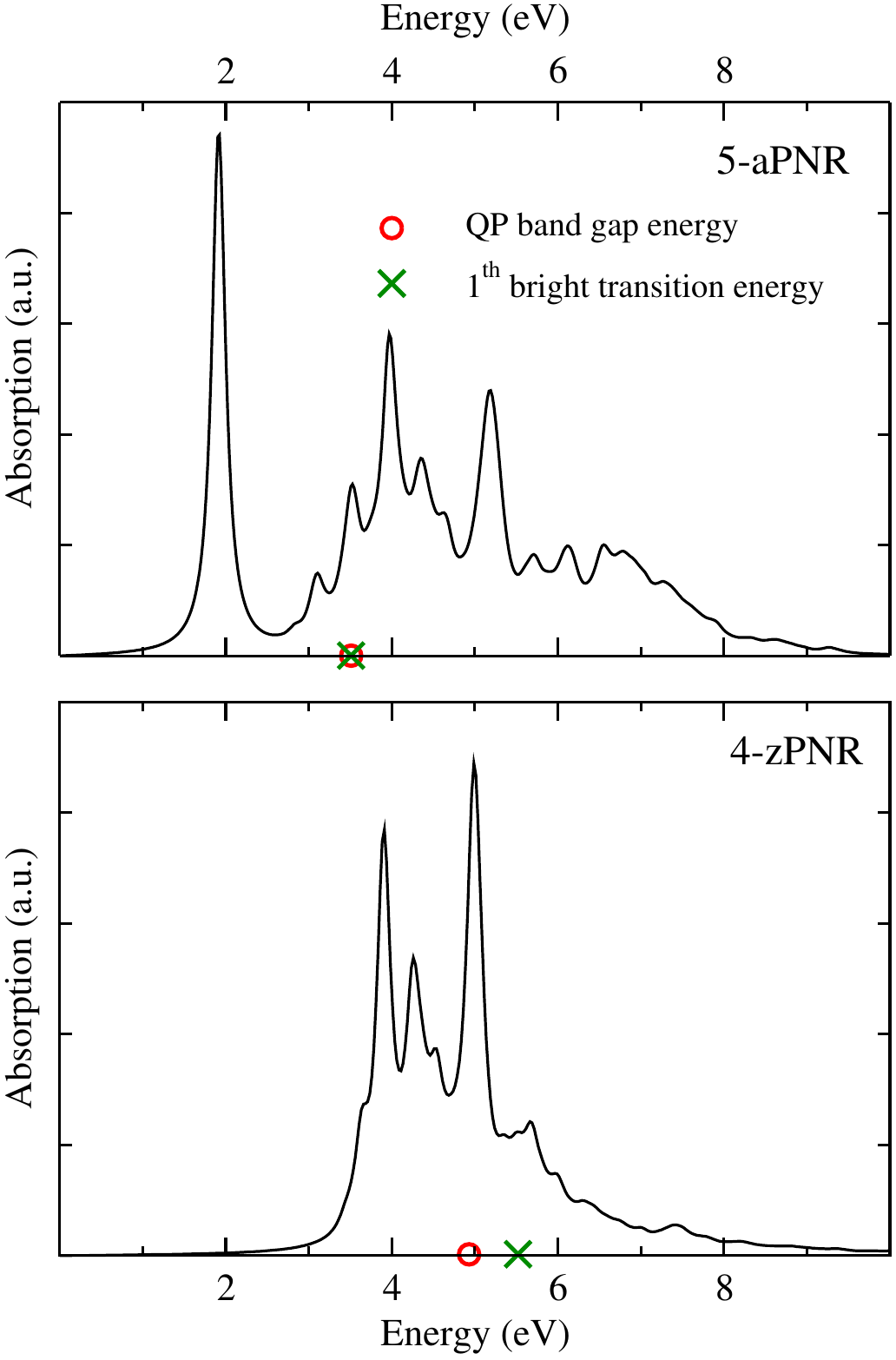}
\caption{\label{fig:abs-ac-zz}
(Color online) Optical absorption spectra of the 5-aPNR, and 4-zPNR.
In nanoribbons, Bethe-Salpeter Hamiltonian are calculated with 10 valence and 15 conduction bands.
The spectra are converged in the displayed energy limit.
}
\end{figure}

\begin{table*}[bht]
 \caption{\label{table:transition}
The states active in the first bright transition of the 5-aPNR, 4-zPNR and phosphorene (with light polarized along the $ac$ and $zz$)
and their projected density of states (PDOS). The percent contribution from each atomic orbital are listed.
HVB (LCB) denotes to Highest Valence Band (Lowest Conduction Band).
The last column shows the QP and optical band gap of these transitions.
 }
\begin{center}
\begin{tabular}{ c | c   c  c c c c c c c c c | c }
 \hline
System &\multicolumn{2}{c}{First Bright Transition}&
~s~ & ~$p_x$~ & ~$p_y$~ & ~$p_z$~ & ~$d_{x^2-y^2}$~ & ~$d_{xy}$~ & ~$d_{zx}$~ & ~$d_{zy}$~ & ~$d_{z^2}$~&Gap~(eV) \\
 ${\vec{e}}$ & ~~$\vec{k}$~~& ~~states~~  &     &     &     &     &     &    &   &       &  & QP/Opt \\
\hline
&&&&&&&&&&&&\\
5-aPNRs &$\Gamma$ &HVB &~$8^{^{(I)}}$ &0 & 0 &$80^{^{(II)}}$&$4^{^{(III)}}$  & 0 &$7^{^{(IV)}}$& 1 &0  & 3.6 \\
  $ac~(x)$ &   &LCB & 4 & ~$15^{^{(I,III)}}$ & 0& $54^{^{(IV)}}$& 0 &0  &$7^{^{(II)}}$ &4& 16 & 1.92 \\
&&&&&&&&&&&&\\&&&&&&&&&&&&\\
phosphorene       &  $\Gamma$ &HVB & $4^{^{(I)}}$& $4^{^{(II)}}$ &0& $73^{^{(III)}}$&$11^{^{(IV)}}$ &0  &$8^{^{(V)}}$&0&0 & 2.2 \\
         $ac~(x)$ &        & LCB& $2^{^{(II)}}$  & $13^{^{(I,IV)}}$&0&$60^{^{(V)}}$ &$3^{^{(II)}}$ &0 & ~$8^{^{(III)}}$ & 1 & $13^{^{(II)}}$ & 1.42\\
&&&&&&&&&&&&\\
\hline
&&&&&&&&&&&&\\
4-zPNRs  & Y &HVB   &	$8^{^{(I)}} $     & $14^{^{(II)}}$  &$10^{^{(III)}}$ &$60^{^{(IV)}}$ &0          &$3^{^{(V)}}$   &0 & $1^{^{(VI)}}$   &~~$4^{^{(VII)}}$ &5.52 \\
     $zz~(y)$ &   &LCB+1&	$11^{^{(III)}} $ &$ 1^{^{(V)}}$ &$22^{^{(I,VII)}}$  &$16^{^{(VI)}}$ &~~$9^{^{(III)}}$ &$27^{^{(II)}}$  &12 &$2^{^{(IV)}}$  & 0 & 3.95 \\
&&&&&&&&&&&&\\&&&&&&&&&&&&\\
phosphorene      &  $(0, Y/4)$ &HVB  &	~~$7^{^{(I)}} $&~~~$2^{^{(II)}}$ &0  &~~~$80^{^{(III)}}$ &~~~$5^{^{(IV)}} $&0 & 5 &0&$1^{^{(V)}}$& 3.53\\
        $zz~(y)$ &          &LCB+1&	4 &0 &~~~~~$65^{^{(I,IV,V)}} $ &0 &2 &$20^{^{(II)}}$&4 &$3^{^{(III)}} $ &2 & 2.93 \\
&&&&&&&&&&&&\\
\hline
\end{tabular}
\end{center}
\end{table*}

\begin{figure*}[ht]
\centering
\includegraphics[width=0.7\linewidth] {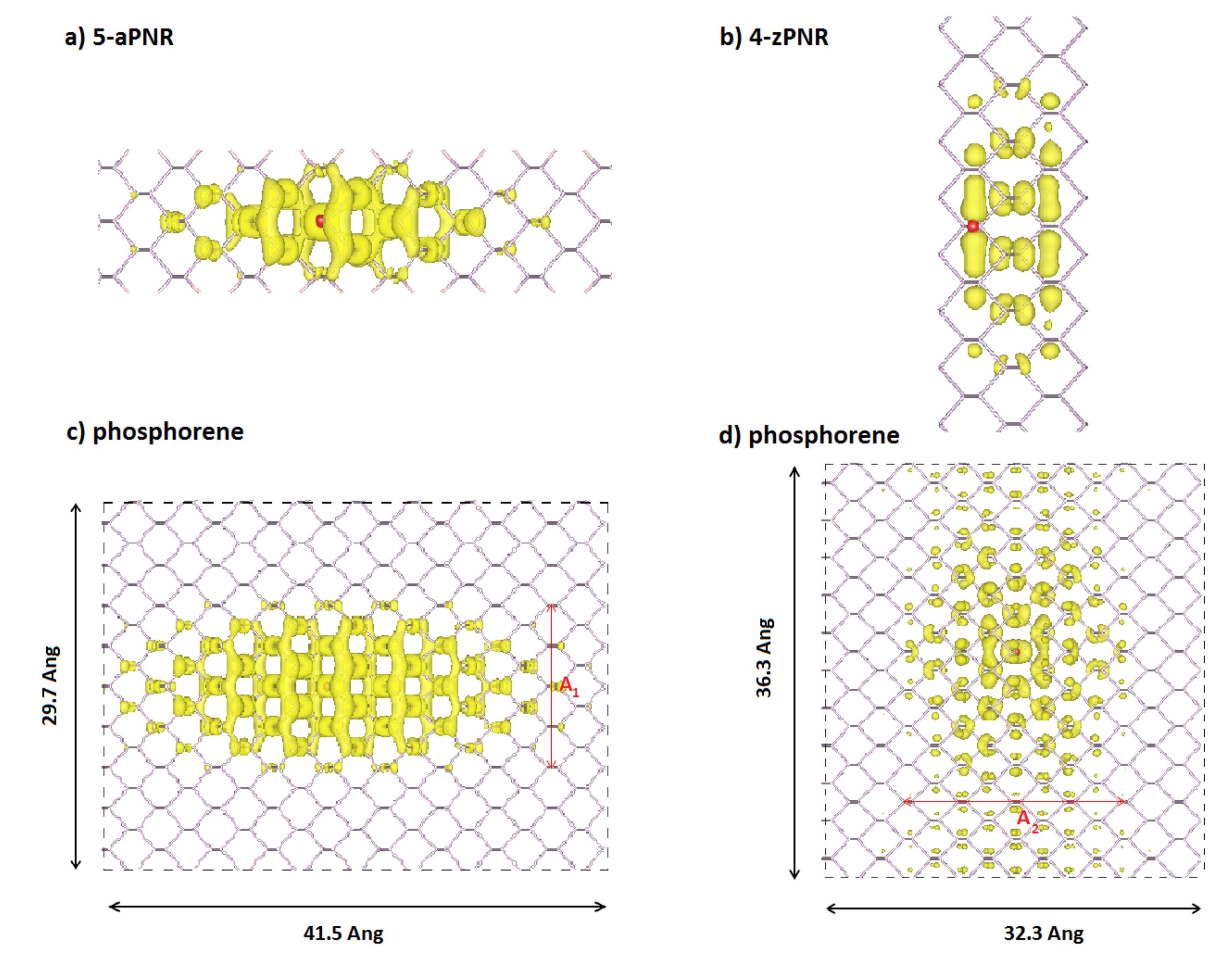}
\caption{\label{fig:excwave} (Color online)
The top view of the electron density distribution of the first bright transition when the hole is fixed at the red point for the (a) 5-aPNR , (b) 4-zPNR , (c) and (d) 2D phosphorene structures. The $\vec{e}$ is along the armchair direction in (a) and (c), however, it is along the zigzag direction in (b) and (d). The hole is located at a high density area of the related states.
The isosurface level of all the plotted density distributions is $0.15~e/\AA^3$.
 }
\end{figure*}

In Fig. \ref{fig:abs-ac-zz}, we plot the optical absorption spectra of the 5-aPNR and 4-zPNR, as an example of aPNRs and zPNRs. The strong excitonic effect is clear in both spectra. There are many invisible (dark) excitonic states around the bright ones which could be active in two photon absorption processes or etc.

The band gap in phosphorus and hydrogenated-PNRs are direct and located at $\Gamma$ point. In aPNRs, the first absorption peak, which is also the dominated absorption peak of system with the energy in the visible light limit arises from the transition between the VBM and CBM at $\Gamma$ point, while optical transitions around the band gap are inactive for all zPNRs. Such a behavior has been seen in 2D phosphorene itself, when the incident light polarization is along the $ac$ or $zz$ direction. In continue, we try to explain the reason of this behavior in detail.

The absorption coefficient which is related to the imaginary part of the dielectric function, $\varepsilon_i(\omega)$, for one photon absorption can be described as
\begin{equation}
 \alpha(\omega)\propto  \varepsilon_i(q  \rightarrow 0,\omega)\propto
  \Sigma_{vc\vec{k}} |\vec{e}\cdot\vec{M}_{cv}(\vec{k}) |^2 \delta(E_{c\vec{k}}-E_{v\vec{k}}-\hbar \omega)
 \end{equation}
  where $\vec{e}$ shows the direction of the incident light polarization, and
 $\vec{M}_{cv}(\vec{k})=\langle v \vec{k} | \vec{p} | c\vec{k} \rangle$ is the dipole matrix element \cite{grosso}.
 The peak of the absorption usually happens in the Van-Hove singularities of joint density of states
 $J_{cv}(\omega)= \int_{B.Z.}\frac{dk}{(2\pi)^3}\delta(E_{c\vec{k}}-E_{v\vec{k}}-\hbar \omega)$
 where their dipole transitions are allowed.

In one photon spectroscopy, the dipole selection rules have two conditions might be satisfied in order to allow transitions; the change in angular momentum between the valence and conduction states should be one ($\Delta l=\pm1$) and since the parity of momentum is odd, the conduction and valence bands should have opposite parity in $\vec{e}$ direction and the same parity in other directions.

Table \ref{table:transition} shows the first bright transitions in the 5-aPNR, 4-zPNR, and phosphorene (as an infinity width limit of PNRs). In aPNRs (zPNRs) the incident light polarization is along the $ac$ ($zz$) direction. In phosphorene, we consider both of the in-plane $ac$ and $zz$ alignment of the light polarization.   Also the projection of the related states into atomic orbital are reported in the Table. The allowed transition between states are specified with the same mark. For the light polarized along the $ac (x)$, the dipole transitions are allowed for $p_x\leftrightarrow (s, d_{x^2-y^2}, d_{z^2}) $ and $p_{y(z)} \leftrightarrow d_{xy(zx)}$ furthermore, the allowed transitions for the light polarized along the $zz (y)$ are $p_y\leftrightarrow (s, d_{x^2-y^2}, d_{z^2}) $ and $p_{x(z)} \leftrightarrow d_{xy(zy)}$ simply by substituting $x$ by $y$.

In the 5-aPNR and phosphorene, with light polarization along the $ac$ direction,  the first bright excitonic bound states originates from the VBM to CBM transition at $\Gamma$ point.  On the contrary, in the 4-zPNR and phosphorene, with light polarization along the $zz$ direction, the gap transition is forbidden and the first absorption peak arises from the transition between the highest valence band (HVB) and the lowest conduction band (LCB)+1 states, in a wave vector away from the $\Gamma$ point. In Fig. \ref{fig:abs-ac-zz}, the second peak of the 4-zPNR with a higher strength comes from transition between the HVB and LCB+3 states at $\Gamma$ point.

Furthermore, Table \ref{table:transition} shows the projected wave functions of the VBM and CBM states of phosphorene at $\Gamma$ point.  According to these data, the occupation of the atomic orbitals of phosphorene are very anisotropic at $\Gamma$ point. This projection also shows why the gap absorption is inactive in phosphorene for light polarization along the $zz$ direction.

Fig. \ref{fig:excwave} presents the distributions of the electron density  for the first bright excitonic states which are specified in table \ref{table:transition}.  The hole is located and fixed at the high electron density area of the related state shown with the red point in the figure.  In the 4-zPNR, the hole position is close to the edges.  We plot the isosurface level of $0.15$~e/\AA.  In the 5-aPNR and 4-zPNR with a high EBE of $\sim 1.5-1.7$~eV, the overlap between the electron and hole wave functions are strong  and excitonic radiuses are about several unit cells along the periodic direction.

In 2D phosphorene when light polarized along the $ac$ direction, the distribution of the electron density is plotted in Fig. \ref{fig:excwave}c.  Because of the high anisotropic effective masses at $\Gamma$ point, the electronic distribution is like an ellipse around the hole and the length along the $ac$ direction with the binding energy of $\sim 0.8$ eV.

Finally, Fig. \ref{fig:excwave}d shows the first bright excitonic state of phosphorene of the light polarized along the $zz$ direction.  This distribution comes from the LCB+1 state at $(0, Y/4)$ with high participation of orbitals along the $y$ ($zz$) direction,  so its distribution along the $zz$ direction is stronger.

In Fig. \ref{fig:excwave}, the smaller length of $A_1$ in comparing with $A_2$ is another evidence in order to have a stronger QC in zPNRs where  these lengths refer to the excitonic Bohr radius of phosphorene.

\section{Summary}\label{sec:conclusion}

We have employed a first-principles GW-BSE simulations to investigate the optical properties of phosphorene nanoribbons (PNR) with armchair (aPNRs) and zigzag (zPNRs) shaped edges.The edge dangling bonds are passivated using hydrogen atoms.

 All the hydrogen-terminated PNRs have a direct band gap at $\Gamma$ point.  However, band structure dispersions of the aPNRs and zPNRs are completely different.  In aPNRs, the valence and conduction band show relativistic dispersion around the gap,  while they are very flat in the zPNRs. From this point of view, zPNRs behaves like a quasi zero-dimensional~\cite{bright} system with stronger quantum confinement effects in compare with aPNRs.

 The quasiparticle band gap, optical gap, and fine structure of excitons
 (excitonic exchange splitting and exciton binding energy) are calculated
 and their tunability by ribbon width are shown separately for both aPNRs and zPNRs. We have obtained that the bang gap behaviors like $w^{-2}$ in an armchair structure where $w$ is the width of a nanoribbon for different DFT approaches, while its exponent changes in a zigzag structure in terms of the different approaches.

We have discussed on the anisotropic behaviors between the aPNRs and zPNRs. Our numerical calculations reveal that the edge states have an important role in the zPNRs behavior and occupancy of the quantum confinement in these nanoribbons, while in the aPNRs, edge states occupied deeper energy states and do not have any participation in the states near the gap. For instance, we have shown that the exchange splitting is quite large and increases by decreasing the size and it is stronger for zPNRs and because of the repulsive exchange interaction term the energy of the singlet state exciton is always higher than the triplet one. Compared to graphene and MoS$_2$ nanoribbons, PNR shows certain advantages.

We have shown that many-body interactions have important impacts on the optical absorption spectra of PNRs and they are not ignorable. In the last part, we have analyzed the first bright excitonic state of PNRs.  The gap transitions are allowed in aPNRs, while in zPNRs the flat band does not satisfy the dipole selection rules of the transition and it is not active in the optical absorption.

\section*{Acknowledgments}

The computational calculations are provided by the high performance computing center of the Institute for Research in Fundamental Sciences (IPM) and Isfahan university of technology (IUT). This work is partially supported by Iran Science Elites Federation grant.

\end {document}